\documentclass[11pt]{article}
\usepackage{jheppub}
\usepackage{tikz,subcaption}
\usetikzlibrary{decorations.pathmorphing}
\tikzset{snake it/.style={decorate, decoration=snake}}
\usetikzlibrary{decorations,decorations.pathmorphing}
\allowdisplaybreaks[1]

\newcommand{\dd}{\mathrm{d}}

\title{Conformal Manifolds with Boundaries or Defects}

\author[a]{Andreas Karch}
\author[b]{and Yoshiki Sato}
\affiliation[a]{Department of Physics, University of Washington, Seattle, Wa, 98195-1560, USA}
\affiliation[b]{Department of Physics, Faculty of Science, The University of Tokyo, Bunkyo-ku, Tokyo 113-0033, Japan }

\preprint{\today}

\abstract{We discuss conformal manifolds for conformal field theories with boundaries or defects.
Using conformal perturbation theory we derive constraints on coefficients appearing in the boundary operator product expansion and three-point functions that need to be satisfied for the existence of marginal couplings.
We present several explicit examples where we confirm that $\beta$-functions vanish using a position space regularization, differential regularization. Where possible, we confirm that our $\beta$-function results agree with the existing literature.
}

\begin{document}
\maketitle


\section{Introduction}

In superconformal field theories, it has been known that there exist families of conformal field theories (CFTs) connected by exactly marginal deformations \cite{Leigh}. Such families are called `conformal manifolds'.
Recently, Refs.~\cite{cm1,cm2} discussed whether conformal manifolds exist in the case without supersymmetry (See also \cite{cm3,cm4}).
Let us consider a candidate marginal deformation of a given CFT\,
\begin{equation}
S_{\mathrm{CFT}}\to S_{\mathrm{CFT}}+g\int \! \dd^dX \, \mathcal{O} (X)
\end{equation}
where $\mathcal{O}$ is a marginal operator whose conformal dimension is $d$.
By using conformal perturbation theory \cite{cpt1}, the $\beta$-function can be determined perturbatively and the order $g^2$ and $g^3$ terms are related with certain three- and four-point functions of the candidate marginal operator respectively.
From the condition that these terms vanish, one obtains a constraint on the coefficients of three-point functions with at least one marginal operator and a sum rule in terms of coefficients of three-point functions and conformal blocks.
If a CFT satisfies the above constraint and the sum rule, there is a possibility that the theory has a conformal manifold.

What happens if boundaries or defects exist?
Naively, we expect that the boundaries or defects do not change the $\beta$-function of an ambient\footnote{In papers discussing defect and boundary conformal field theories in the context of holography, it has been common to use the word ``ambient" space for the spacetime of the field theory away from the boundary so that the word ``bulk" can be reserved for the spacetime of the holographic dual. While this current paper does discuss holography only tangentially, we still stick with this convention in order to avoid any future confusion.} operator since a boundary marginal operator is expected not to influence correlation functions of ambient operators away from the boundary. Correspondingly the $\beta$-function of ambient operators does not give new interesting information.
In contrast, the ambient marginal operator does drive the $\beta$-function of the boundary operator.
In two-dimensional CFTs, these aspects were explored in \cite{cm0}.
In this paper, we generalize the results in \cite{cm1,cm2} to CFTs with boundaries or defects.
Or we can also say that we generalize the results in \cite{cm0} to higher dimensional CFTs.

This paper is organized as follows.
In section \ref{sec2}, we will derive necessary conditions such that conformal manifolds with boundaries or defects exist by using conformal perturbation theory.
In section \ref{sec3}, we will give simple examples of conformal manifolds.
To exhibit that $\beta$-functions of these examples vanish, we employ differential regularization.
We will conclude in section \ref{sec4}.
In appendix \ref{app}, we give a detailed discussion of the asymptotic behavior of conformal blocks.

\section{Constraints from conformal perturbation theory}
\label{sec2}

Let us consider boundary conformal field theories (BCFTs) or defect conformal field theories (dCFTs) and deform them by introducing an marginal ambient operator $\mathcal{O}$ and a marginal boundary operator $\mathcal{\hat{O}}$,
\begin{equation}
S_{\text{BCFT}} \to S_{\text{BCFT}}+g \int \! \dd^dX \, \mathcal{O} (x,w) + \hat{g}\int \! \dd ^{d-1}x \, \mathcal{\hat{O}} (x) \,.
\end{equation}
The conformal dimension of the ambient operator is $\Delta=d$, and that of  the boundary operator is $\hat{\Delta}=d-1$.
We use capital letters for the ambient coordinates and lower case letters for the boundary coordinates.
$w$ represents a coordinate perpendicular to the boundary or defect.
We use hats for quantities on the boundary.
The $\beta$-functions of these operators are obtained by using conformal perturbation theory \cite{cpt1}.
Following \cite{cpt2}, we consider the overlaps
\begin{equation}
\langle \mathcal{O} (\infty) |0 \rangle_{g,\hat{g},V,\hat{V}}
\end{equation}
and
\begin{equation}
\langle \mathcal{\hat{O}} (\infty) |0 \rangle_{g,\hat{g},V,\hat{V}}
\end{equation}
where $\mathcal{O}(\infty) =\lim_{X\to \infty} X^{2d} \mathcal{O}(X) $ and $\mathcal{\hat{O}}(\infty) =\lim_{x\to \infty} x^{2(d-1)} \mathcal{\hat{O}}(x) $.
The state
\begin{equation}
 |0 \rangle_{g,\hat{g},V,\hat{V}} = \exp \left( g \int _V   \mathcal{O} + \hat{g}\int_{\hat{V}}  \mathcal{\hat{O}} \right) |0 \rangle
\end{equation}
is obtained by deforming the theory in a finite region surrounding the origin.
Note that $V$ ($\hat{V}$) is a small region surrounding the origin in the ambient (boundary) spacetime.
Here we introduce $\int _V = \int _V  \dd ^d X$ and $\int _{\hat{V}} =\int_{\hat{V}} \dd ^{d-1}x$ to simplify expressions.
We can obtain the $\beta$-functions by demanding that these overlaps do not depend on an UV cut-off scale.
By expanding the overlaps in terms of coupling constants, we obtain
\begin{equation}
\begin{split}
\langle \mathcal{O} (\infty) |0 \rangle_{g,\hat{g},V,\hat{V}} & \simeq
\langle \mathcal{O} (\infty) \rangle + \hat{g}\int_{\hat{V}} \langle \mathcal{O}(\infty) \mathcal{\hat{O}} (x) \rangle +\frac{\hat{g}^2}{2}\int_{\hat{V}_1} \int_{\hat{V}_2} \langle \mathcal{O}(\infty) \mathcal{\hat{O}} (x_1) \mathcal{\hat{O}} (x_2)\rangle  \\
&+\frac{\hat{g}^3}{6}\int_{\hat{V}_1} \int_{\hat{V}_2}  \int_{\hat{V}_3} \langle \mathcal{O}(\infty) \mathcal{\hat{O}} (x_1) \mathcal{\hat{O}} (x_2) \mathcal{\hat{O}} (x_3) \rangle  \\
&+g \int_{V}  \langle \mathcal{O}(\infty) \mathcal{O} (X) \rangle +\frac{g^2}{2} \int_{V_1}  \int_{V_2}  \langle \mathcal{O}(\infty) \mathcal{O} (X_1)\mathcal{O} (X_2) \rangle  \\
&+g\hat{g} \int_{V}  \int_{\hat{V}}  \langle \mathcal{O}(\infty) \mathcal{O} (X)\mathcal{\hat{O}}(x) \rangle  \\
&+\frac{g^2\hat{g} }{2} \int_{V_1} \int_{V_2}  \int_{\hat{V}} \langle \mathcal{O}(\infty) \mathcal{O} (X_1)\mathcal{O} (X_2)\mathcal{\hat{O}}(x) \rangle \\
&+\frac{g\hat{g}^2 }{2} \int_{V} \int_{\hat{V}_1}  \int_{\hat{V}_2} \langle \mathcal{O}(\infty) \mathcal{O} (X)\mathcal{\hat{O}} (x_1)\mathcal{\hat{O}}(x_2) \rangle
+\cdots \,,
\end{split}
\label{bulk-op}
\end{equation}
and
\begin{equation}
\begin{split}
\langle \mathcal{\hat{O}} (\infty) |0 \rangle_{g,\hat{g},V,\hat{V}} &\simeq
\langle \mathcal{\hat{O}} (\infty) \rangle +\hat{g}\int_{\hat{V}}  \langle \mathcal{\hat{O}}(\infty) \mathcal{\hat{O}} (x) \rangle +\frac{\hat{g}^2}{2}\int_{\hat{V}_1}  \int_{\hat{V}_2} \langle \mathcal{\hat{O}}(\infty) \mathcal{\hat{O}} (x_1) \mathcal{\hat{O}} (x_2)\rangle  \\
&+\frac{\hat{g}^3}{6}\int_{\hat{V}}  \int_{\hat{V}_1} \int_{\hat{V}_2} \langle \mathcal{\hat{O}}(\infty) \mathcal{\hat{O}} (x_1) \mathcal{\hat{O}} (x_2) \mathcal{\hat{O}} (x_3) \rangle  \\
&+g \int_{V} \langle \mathcal{\hat{O}}(\infty) \mathcal{O} (X) \rangle +\frac{g^2 }{2} \int_{V}  \int_{V}  \langle \mathcal{\hat{O}}(\infty) \mathcal{O} (X_1)\mathcal{O} (X_2) \rangle  \\
&+g\hat{g} \int_{V} \int_{\hat{V}} \langle \mathcal{\hat{O}}(\infty) \mathcal{O} (X)\mathcal{\hat{O}}(x) \rangle  \\
&+\frac{g^2\hat{g} }{2} \int_{V_1}  \int_{V_2} \int_{\hat{V}}  \langle \mathcal{\hat{O}}(\infty) \mathcal{O} (X_1)\mathcal{O} (X_2)\mathcal{\hat{O}}(x) \rangle \\
&+\frac{g\hat{g}^2 }{2} \int_{V}  \int_{\hat{V}_1} \int_{\hat{V}_2}  \langle \mathcal{\hat{O}}(\infty) \mathcal{O} (X)\mathcal{\hat{O}} (x_1)\mathcal{\hat{O}}(x_2) \rangle
+\cdots \,,
\end{split}
\label{boundary-op}
\end{equation}
where the dots represent higher order terms.
To obtain the $\beta$-functions, we have to pick up logarithmic divergent parts in \eqref{bulk-op} and \eqref{boundary-op}.
The $\beta$-function of the ambient operator is not expected to give any new constraints because the boundary operator does not change the ambient theory.
In fact, at leading order of $\hat{g}$, the two-point function is involved and it does not diverge.
At the next order, one can show that the three-point function does not diverge logarithmically using a conformal block decomposition \eqref{block} which we will discuss later. At the order $g\hat{g}$, it is almost impossible to evaluate the contribution explicitly because this involves a three-point function with two ambient operators and one boundary operator.
However, if we assume the order between two coupling constants as $g\sim \hat{g}^2$, we can consider this term as a higher order term.
We will discuss the order between $g$ and $\hat{g}$ and difficulties of computations of three-point functions later.
In total, from the ambient $\beta$-function, we get the same constraints as obtained in \cite{cm1,cm2}.
We do not repeat their computations and do not write the constraints explicitly, here.

Hence our main attention is the overlap of the boundary operator \eqref{boundary-op}.
The first term is the one-point function and does not give any logarithmic divergence.
The next three terms are the counterparts of similar terms in the $\beta$-function of the ambient operator.
From them we get constraints similar to those of the ambient operator case except for the fact that the corresponding correlation functions live entirely on the boundary.
The last five terms are peculiar to BCFTs or dCFTs and are the interesting terms we wish to analyze.

Before evaluating each term, we comment on the relative order between $g$ and $\hat{g}$.
In general, they can be dialed independently and we mainly assume $g\sim \hat{g}^2$ throughout this paper except one example encountered in section \ref{sec31}.
One important rationale for this choice is the fact that the ambient operator can be regarded as two boundary operators when using the method of images to enforce the boundary conditions. Using this standard method the insertion of a single ambient operator together with its mirror operator for the purposes of actually evaluating the correlators effectively inserts two operators into the correlation function. So if $\hat{g}$ counts the number of boundary operators inserted, in terms of determining the difficulty of the calculation to be performed it makes sense to count $g \sim \hat{g}^2$. With this scaling we can also argue that the very complicated computation of the order $g^2$-term, which we will discuss later, is higher order than the terms we discuss in detail and can be neglected for our purposes.

\subsection*{Order $g$-term}
The two-point function of an ambient operator and a boundary operator is determined as \cite{McAvity}
\begin{equation}
\langle \mathcal{O}(x_1,w) \mathcal{\hat{O}}(x_2) \rangle
=\frac{B_{\mathcal{O}\mathcal{\hat{O}} }}{(2w)^{\Delta-\hat{\Delta}}(x_{12}^2+w^2  )^{\hat{\Delta}} }
\end{equation}
where $x_{ij}$ means $x_i-x_j$ and $B_{\mathcal{O}\mathcal{\hat{O}}} $ is a coefficient appearing in the boundary operator product expansion,
\begin{equation}
\mathcal{O}(x,w)=\sum_n \frac{B_{\mathcal{O}\mathcal{\hat{O}}_n}} {(2w)^{\Delta-\hat{\Delta}_n}}\mathcal{\hat{O}}_n (x)\,.
\label{bope}
\end{equation}
Then the order $g$-term in \eqref{boundary-op} becomes
\begin{equation}
g \int_{V} \! \dd^{d}X \langle \mathcal{\hat{O}}(\infty) \mathcal{O} (X) \rangle
=g \int_{V} \! \dd^{d}X \, \frac{B_{\mathcal{O}\mathcal{\hat{O}} }}{(2w)^{\Delta-\hat{\Delta}} }
\end{equation}
where we normalize two-point functions such that $\langle \mathcal{\hat{O}}(\infty)\mathcal{\hat{O}}(0) \rangle =1 $.
The integral about $x$ gives only a volume factor\footnote{In our notation, $X=(x,w)$.}.
A logarithmic divergence comes from the integral about $w$. Since $\Delta -\hat{\Delta}=1$ we can evaluate
\begin{equation}
g \int_{V} \! \dd^{d}X \, \frac{B_{\mathcal{O}\mathcal{\hat{O}} }}{(2w)^{\Delta-\hat{\Delta}} } \sim g  B_{\mathcal{O} \mathcal{\hat{O}}} \log \Lambda
\end{equation}
 where we ignore unimportant prefactors to determine necessary conditions such that the $\beta$-function vanishes. Hence the necessary condition is
\begin{equation}
B_{\mathcal{O} \mathcal{\hat{O}}}=0
\end{equation}
for the marginal operators $\mathcal{O}$ and $\mathcal{\hat{O}}$.

\subsection*{Order $g\hat{g}$-term}

Next, we evaluate the divergence of the order $g\hat{g}$-term.
To the best of our knowledge, the conformal block decomposition of a three-point function with one ambient operator and two boundary operators has not been obtained before except in two-dimensional CFTs \cite{Lewellen}.
Hence we give a detailed derivation of the conformal block decompositions by using a Casimir method here.

A three-point function with one ambient operator and two boundary operators has the following general form,
\begin{equation}
\begin{split}
&\langle \mathcal{O}_1 (x_1,w) \mathcal{\hat{O}}_2(x_2)\mathcal{\hat{O}}_3(x_3) \rangle \\
&=(-2P_1\cdot \hat{P}_2)^{-\frac{\Delta_1+\hat{\Delta}_2-\hat{\Delta}_3}{2}}
(-2P_1\cdot \hat{P}_3)^{-\frac{\Delta_1+\hat{\Delta}_3-\hat{\Delta}_2}{2}}
(-2\hat{P}_2\cdot \hat{P}_3)^{-\frac{\hat{\Delta}_2+\hat{\Delta}_3-\Delta_1}{2}} f(\eta) \,.
\end{split}
\end{equation}
Here we introduce embedding coordinates $P_1,\hat{P}_2,\hat{P}_3$ and $B$, and they are expressed as
\begin{equation}
\begin{split}
P_1& =(1,x_1^2+w^2,x_1,w) \,, \qquad  \hat{P}_2 =(1,x_2^2,x_2,0) \,, \\
\hat{P}_3& =(1,x_3^2,x_3,0) \,, \qquad B=(0,0,0,1)
\end{split}
\end{equation}
on the projective null cone and $\eta$ is the conformal cross-ratio,
\begin{equation}
\eta = \frac{ (-2P_1\cdot \hat{P}_2) (-2P_1\cdot \hat{P}_3) }{ (-2\hat{P}_2\cdot \hat{P}_3) (P_1 \cdot B)^2}
= \frac{(x_{12}^2+w^2)(x_{13}^2+w^2) }{x_{23}^2 w^2}  \,.
\end{equation}
 See \cite{Liendo} for the details of the embedding formalism in BCFTs and also for the Casimir method for the conformal block decomposition of two-point functions.
The three-point function satisfies the Casimir differential equation,
\begin{equation}
\hat{L}_1^2 \langle \mathcal{O}_1 \mathcal{\hat{O}}_2 \mathcal{\hat{O}}_3 \rangle
=-C_{\hat{\Delta}} \langle \mathcal{O}_1 \mathcal{\hat{O}}_2 \mathcal{\hat{O}}_3 \rangle
\end{equation}
with $C_{\hat{\Delta}}=\hat{\Delta}(\hat{\Delta}-d+1)$.
The differential operator $\hat{L}$ is defined as
\begin{equation}
\hat{L}_{\hat{A}\hat{B}}:=P_{\hat{A}}\frac{\partial}{\partial P^{\hat{B}}} - P_{\hat{B}}\frac{\partial}{\partial P^{\hat{A}}}
\end{equation}
and its square is given by
\begin{align}
\hat{L}^2&:=\frac{1}{2} \hat{L}_{\hat{A}\hat{B}} \hat{L}^{\hat{A}\hat{B}} \notag \\
&=P_{\hat{A}} P^{\hat{A}} \frac{\partial}{\partial P^{\hat{B}}} \frac{\partial}{\partial P_{\hat{B}}} -P_{\hat{A}} \frac{\partial}{\partial P_{\hat{A}}} \left( P_{\hat{B}} \frac{\partial}{\partial P_{\hat{B}}} \right) -(d-1) P_{\hat{A}} \frac{\partial}{\partial P_{\hat{A}}}
\end{align}
Note that $\hat{A}$ runs from $-,+,1,\cdots,d-1$.
The Casimir differential equation reduces to a second-order ordinary differential equation
\begin{equation}
\begin{split}
&\left( \frac{4\alpha \beta}{\eta} -(\alpha+\beta)^2-(d-1)(\alpha+\beta)+\hat{\Delta}(\hat{\Delta}-d+1) \right) f(\eta)  \\
&+(4(\alpha+\beta+1)(1-\eta)-2(d-1)\eta)f'(\eta)+4\eta (1-\eta)f''(\eta) =0
\end{split}
\end{equation}
where we introduced
\begin{equation}
\alpha = -\frac{\Delta_1+\hat{\Delta}_2-\hat{\Delta}_3}{2} \,, \qquad \beta =-\frac{\Delta_1+\hat{\Delta}_3-\hat{\Delta}_2}{2}
\end{equation}
to simplify expressions.
Solutions of the Casimir equation are given by
\begin{equation}
\begin{split}
f(\eta)&= C_1 \eta^{-\alpha} {}_2F_1 \left( \frac{d-1-\alpha+\beta-\hat{\Delta}}{2},\frac{-\alpha+\beta+\hat{\Delta}}{2},1-\alpha+\beta ,\eta \right)  \\
&+ C_2 \eta^{-\beta} {}_2F_1 \left( \frac{d-1+\alpha-\beta-\hat{\Delta}}{2},\frac{\alpha-\beta+\hat{\Delta}}{2},1+\alpha-\beta ,\eta \right)
\end{split}
\end{equation}
where $C_1$ and $C_2$ are constants of integration.
Since the three-point function is symmetric under exchange of $\hat{O}_2$ and $\hat{O}_3$,
$f(\eta)$ should be symmetric in a similar manner.
Thus, we choose $C_1=C_2$.
We can determine the integration constants explicitly from the asymptotic behavior of $f(\eta)$.
However, here we do not need to fix it since the overall normalization is not important for our purpose of determining when the associated divergences vanish. See appendix \ref{app} for the detail of the asymptotic behavior of $f(\eta)$.
Then, the three-point function can be decomposed as,
\begin{equation}
\langle \mathcal{O}_1(x_1,w) \mathcal{\hat{O}}_2(x_2) \mathcal{\hat{O}}_3(x_3) \rangle = \sum_\ell \frac{
B_{\mathcal{O}_1\mathcal{\hat{O}}_\ell}
\hat{C}_{\mathcal{\hat{O}}_\ell \mathcal{\hat{O}}_2 \mathcal{\hat{O}}_3 } f_\ell(\eta)
}{
(x_{12}^2+w^2)^{\frac{\Delta_1+\hat{\Delta}_2-\hat{\Delta}_3}{2}}(x_{13}^2+w^2)^{\frac{\Delta_1+\hat{\Delta}_3-\hat{\Delta}_2}{2}} x_{23}^{\frac{\hat{\Delta}_2+\hat{\Delta}_3-\Delta_1}{2}} }
\label{block}
\end{equation}
where the conformal block is given by
\begin{equation}
\begin{split}
f_\ell(\eta)&= C \eta^{\frac{\Delta_1+\hat{\Delta}_2-\hat{\Delta}_3}{2}} {}_2F_1 \left( \frac{d-1+\hat{\Delta}_2-\hat{\Delta}_3 -\hat{\Delta}_\ell}{2},\frac{\hat{\Delta}_2-\hat{\Delta}_3 -\hat{\Delta}_\ell}{2},1+\hat{\Delta}_2-\hat{\Delta}_3 ,\eta \right)  \\
&+ C \eta^{\frac{\Delta_1+\hat{\Delta}_3-\hat{\Delta}_2}{2}} {}_2F_1 \left( \frac{d-1+\hat{\Delta}_3-\hat{\Delta}_2 -\hat{\Delta}_\ell}{2},\frac{\hat{\Delta}_3-\hat{\Delta}_2+\hat{\Delta}_\ell}{2},1+\hat{\Delta}_3-\hat{\Delta}_2 ,\eta \right)
\end{split}
\end{equation}
with an unfixed coefficient $C$.
Here $B_{\mathcal{O}\mathcal{\hat{O}}_\ell}$ is a coefficient appearing in the boundary operator product expansion as before and $\hat{C}_{\mathcal{\hat{O}}_\ell \mathcal{\hat{O}} \mathcal{\hat{O}}}$ is the coefficient appearing in the three-point function of boundary operators,
\begin{equation}
\langle \mathcal{\hat{O}}_1(x_1) \mathcal{\hat{O}}_2(x_2) \mathcal{\hat{O}}_3(x_3) \rangle =\frac{ \hat{C}_{\mathcal{\hat{O}}_1 \mathcal{\hat{O}}_2 \mathcal{\hat{O}}_3 }}{
x_{12}^{\hat{\Delta}_1+\hat{\Delta}_2-\hat{\Delta}_3}x_{23}^{\hat{\Delta}_2+\hat{\Delta}_3-\hat{\Delta}_1} x_{31}^{\hat{\Delta}_3+\hat{\Delta}_1-\hat{\Delta}_2} } \,.
\end{equation}


Let us evaluate a divergence in the three-point function with two identical boundary operators.
Taking the $x_3\to \infty$ limit and taking the normalization into account, the three-point function becomes
\begin{equation}
\langle \mathcal{O}(x_1,w) \mathcal{\hat{O}}(x_2) \mathcal{\hat{O}}(\infty) \rangle = \sum_\ell
\frac{ B_{\mathcal{O}\mathcal{\hat{O}}_\ell} \hat{C}_{\mathcal{\hat{O}_\ell} \mathcal{\hat{O}} \mathcal{\hat{O}} } f_\ell(\eta)}{
(x_{12}^2+w^2)^{\Delta/2} } \,.
\end{equation}
The integration over $x_2$ gives a volume factor and hence the total integral of the three-point function reduces to
\begin{equation}
\int \! \dd w \int \dd^{d-1} x\, (x^2+w^2)^{-\Delta/2} \eta^{\Delta/2} {}_2F_1 \left( \frac{d-1-\hat{\Delta}_\ell}{2},\frac{\hat{\Delta}_\ell}{2},1,\eta \right)
\end{equation}
where $\eta=(x^2+w^2)/w^2$. Note that the integration regions are not the whole of spacetime but the small region surrounding the origin.
After a change of variables $x\to wx$, the integral becomes
\begin{equation}
\int \! \dd w \int \dd r \, w^{d-1-\Delta} r^{d-2} \eta^{\Delta/2} {}_2F_1 \left( \frac{d-1-\hat{\Delta}_\ell}{2},\frac{\hat{\Delta}_\ell}{2},1,\eta \right)
\end{equation}
up to unimportant prefactors. Here $r$ is a radial coordinate. The $r$-integral does not diverge and the $w$-integral diverges logarithmically.
Correspondingly we get a constraint
\begin{equation}
\sum_\ell B_{\mathcal{O}\mathcal{\hat{O}}_\ell} \hat{C}_{\mathcal{\hat{O}_\ell} \mathcal{\hat{O}} \mathcal{\hat{O}} } =0
\end{equation}
for the marginal operators $\mathcal{O}$ and $\mathcal{\hat{O}}$.
The summation over $\ell$ runs over all operators appearing in the boundary operator product expansion of the marginal ambient operator.

\subsection*{Other terms}
When we dial $g$ and $\hat{g}$ independently, we can regard other terms as higher order terms by setting $g \sim \hat{g}^2$ as noted above.
However, they cannot be dialed independently in some examples like super Janus as we will see later in subsection~\ref{sec31}.
In the situation where $g\sim \hat{g}$, we have to evaluate three-point functions with two ambient operators and one boundary operator in order to determine the order $g^2 \hat{g}$ contribution, which in a scheme where $g \sim \hat{g}$ is of the same order as the $\hat{g}^3$ term we accounted for. However, the order $g^2 \hat{g}$ contribution is significantly more complicated than the three-point functions with one ambient operator and two boundary operators we calculated above: the former can be regarded as five-point function when using the folding trick while the latter can be regarded as four-point function.
In practice, when we decompose the three-point functions to conformal blocks, they depend on two conformal cross-ratios and a Casimir differential equation becomes a partial differential equation. Like conformal blocks of a four-point function in standard CFTs, it might be difficult to obtain analytical solutions of this partial differential equation.
Hence we do not further consider evaluating constraints from this three-point function.
It is necessary to check whether new constraints are compatible with other constraints when $g\sim \hat{g}$.
As we will see later, super Janus does not rule out the existence of marginal couplings with this scaling and we hope that this new constraint is consistent with the other constraints.

We can guess the form of the new constraint.
The three-point function can be decomposed to conformal blocks,
\begin{equation}
\langle \mathcal{\hat{O}}(\infty) \mathcal{O}(X_1) \mathcal{O}(X_2)  \rangle
= \sum_{m,n} B_{\mathcal{O}\mathcal{\hat{O}}_m} B_{\mathcal{O}\mathcal{\hat{O}}_n} \hat{C}_{\mathcal{\hat{O}}\mathcal{\hat{O}}_m\mathcal{\hat{O}}_n } F (\eta_1, \eta_2)
\end{equation}
where $\eta_1$ and $\eta_2$ are conformal cross-ratios.
An integration in terms of $X_2$ gives a volume factor and our concern is how the integration in terms of $X_1$ behaves. If it does not diverge at all, it obviously does not give any new constraint.
If it diverges logarithmically, we will get a constraint,
\begin{equation}
 \sum_{m,n} B_{\mathcal{O}\mathcal{\hat{O}}_m} B_{\mathcal{O}\mathcal{\hat{O}}_n} \hat{C}_{\mathcal{\hat{O}}\mathcal{\hat{O}}_m\mathcal{\hat{O}}_n } =0 \,.
\end{equation}
This seems a reasonable constraint.
In all other cases, the constraint would be more complicated.

\section{Examples}
\label{sec3}

\subsection{Super Janus}
\label{sec31}

One candidate for conformal manifolds in a CFT with boundaries or defects is super Janus \cite{sJanus1,sJanus2}.
Janus type field theories often have holographic duals in terms of the Janus solution of type IIB supergravity \cite{Janus} or related solutions. Their defining characteristic is that the gauge coupling constant jumps across the defect. We can consider this jumping coupling constant as the ambient marginal deformation operator. This becomes more apparent when we employ the folding trick.
Without supersymmetry, such a field theory has no boundary marginal operator added to the action and it is a good candidate for conformal manifolds. Note that in the Janus case we start out with a CFT without defects or boundaries and the defect gets only introduced by the particular ambient space operator we introduce. This means that the constraints we have to check are identical to those in a CFT without boundary. Indeed the marginal operator of the Janus field theory satisfies constraints obtained in \cite{cm1,cm2}. One could argue that Janus is just one more example of the results obtained in \cite{cm1,cm2}, but the fact that the deformation results in a dCFT is somewhat non-trivial. At strong coupling, one can use the holographic dual to see that at least in this regime the Janus deformation is, in fact, exactly marginal to all orders: the dual geometry has an AdS$_4$ factor, which indicates an unbroken conformal invariance, even in the presence of an order one Janus deformation.

To restore supersymmetry in the Janus field theory, we need to add boundary terms and we can regard these boundary terms as boundary marginal operators.
The Lagrangian of the $\mathcal{N}=4$ super Yang-Mills theory is written as
\begin{equation}
\begin{split}
\mathcal{L}_{\mathcal{N}=4} =& -\partial_\mu \phi_I^\ast \partial^\mu \phi_I -\frac{i}{2} \bar{\psi}_I \Gamma^\mu \partial_\mu \psi_I + F_I^\ast F_I +W'_I F_I -\frac{i}{2}W''_{IJ}\bar{\psi}_IP_+\psi_J \\
&-\frac{1}{4g^2}F_{\mu \nu}^a F^{a\, \mu \nu}-\frac{i}{2g^2} \bar{\lambda}^a \Gamma^\mu D_\mu \lambda^a + \frac{1}{2g^2}D^a D^a +\mathcal{L}_{\text{int}}
\end{split}
\end{equation}
where we use different normalizations for the chiral multiplets and the vector multiplets and interaction terms are not written explicitly.
The Lagrangian of super Janus is constructed as
\begin{equation}
\mathcal{L}_{\text{super Janus}}=\mathcal{L}_{\mathcal{N}=4}-\gamma \varepsilon (w) \mathcal{L}_{\mathcal{N}=4} -2\partial_{w} g \, \text{Im} \frac{\delta W}{\delta g}-\partial_{w} \left( \frac{1}{4g^2} \right) \bar{\lambda}^a \Gamma^5 \lambda ^a
\end{equation}
where $\gamma$ is a dimensionless parameter and expressed by using gauge coupling constants $g_+$ for $w>0$ and $g_-$ for $w<0$,
\begin{equation}
\gamma=\frac{g_+^2-g_-^2}{g_+^2+g_-^2} \,.
\end{equation}
From the Lagrangian, the ambient operator is determined as
\begin{equation}
\gamma \mathcal{O}=-\gamma \varepsilon (x_3) \mathcal{L}_{\mathcal{N}=4}
\end{equation}
and two boundary operators are given by
\begin{equation}
\gamma \bar{g} \mathcal{\hat{O}}_1 =-2\gamma \bar{g} \, \text{Im} \frac{\delta W}{\delta g}
\end{equation}
and
\begin{equation}
\frac{\gamma}{\bar{g}^2} \mathcal{\hat{O}}_2 =\frac{\gamma}{2\bar{g}^2}\bar{\lambda}^a \Gamma^5 \lambda^a
\end{equation}
respectively.
From these expressions, we see that it is not possible to dial the two boundary operators independently.
In the situation where $\bar{g}\sim 1$, the ambient operator and the boundary operators are of the same order.
In this case, we have to evaluate a three-point function with two ambient operators as mentioned before.

We can check perturbatively whether super Janus has a conformal manifold by explicit computation of correlation functions.
In fact, all relevant correlation functions vanish trivially and super Janus is indeed a possible candidate of conformal manifolds with boundaries or defects. Once again, the existence of holographically dual supergravity solutions \cite{sJanus2,DHoker:2006vfr,Suh} to super Janus involving an AdS$_4$ factor indicates that at least at strong coupling this theory indeed does have a conformal manifold. Our discussion above applied to minimal supersymmetric Janus, where counter terms were added in order to restore at least some supersymmetry. Janus solutions preserving extended supersymmetries have also been constructed \cite{DHoker:2007zhm} and the dual field theories for these maximally supersymmetric Janus type field theories are of course also candidates for dCFTs with conformal manifolds.

If we do not add any counter terms to restore supersymmetry, we obtain an example of a conformal manifold without supersymmetry. We suspect that if we were to add the counter terms with arbitrary coefficients, in particular with our preferred $g \sim \hat{g}^2$ scaling, we would still obtain a dCFT with a conformal manifold. However any supersymmetry, and any connection to a known holographic dual, would be lost.

\subsection{Mixed dimensional QED}
\label{sec32}
Mixed dimensional quantum electrodynamics (QED) is standard QED coupled to fermions localized on a lower-dimensional boundary or defect. Here we restrict our attention to four-dimensional QED coupled with a fermion on a co-dimension one boundary. This model was explored by \cite{Teber,HH} and we will review it and give a new derivation of the $\beta$-function by using a differential regularization \cite{diff} which is a position space regularization and hence more suited to the task of regulating theories without translation invariance.
As noted in \cite{HH}, this model is conformal to all orders in perturbation theory. Reproducing at least the leading order result is a reassuring check of our methods.

The Lagrangian of mixed QED is given by
\begin{equation}
S=-\frac{1}{4} \int \! \dd ^4 X\, F_{\mu \nu}^2 +\int \! \dd^3 x \, i \bar{\psi} \gamma^i D_i \psi
\end{equation}
where the covariant derivative is $D_i =\partial _i -ig A_i$ and the metric is mostly plus. The ambient space has a boundary at $w=0$ and the fermion is localized on this boundary.
In \cite{HH}, it was shown that the $\beta$-function of the mixed dimensional QED vanishes by using standard momentum space methods. However, boundaries or defects break translation invariance and Fourier transformation to momentum space is challenging except in some simple situations. For instance, the Janus type field theory does not work well in momentum space. Most notably, the propagator which is easily constructed in position space using the method of images, does not have a simple momentum space representation.
Since we would like to consider such Janus type field theory in this paper, we reproduce results in \cite{HH} using a regularization directly in position space.

Here we summarize our notation, basically following \cite{HH}.
A propagator for a $d$-dimensional scalar field is given by
\begin{equation}
\label{janusprop}
G_{\mathrm{S}}(x_1,w_1,x_2,w_2)=C_{\mathrm{S}}\left[ \frac{1}{((x_1-x_2)^2+(w_1-w_2)^2)^{\frac{d-2}{2}}}+\frac{1}{((x_1-x_2)^2+(w_1+w_2)^2)^{\frac{d-2}{2}}}\right]
\end{equation}
with coefficient
\begin{equation}
C_{\mathrm{S}}=\frac{1}{(d-2)\mathrm{Vol}(S^{d-1}) }\,,  \qquad \mathrm{Vol}(S^{d-1})=\frac{2\pi^{d/2}}{\Gamma (d/2)}\,.
\end{equation}
Using this propagator, the propagator of a gauge field can be written as
\begin{equation}
G_{\mathrm{G}}^{\mu \nu}(x_1,w_1,x_2,w_2)=\eta^{\mu \nu}G_{\mathrm{S}}(x_1,w_1,x_2,w_2)\,.
\end{equation}
A propagator of a fermion on co-dimension one boundary is given by
\begin{equation}
G_{\mathrm{F}} (x_1,x_2)=C_{\mathrm{F}} \frac{\gamma^i(x_1-x_2)_i}{((x_1-x_2)^2)^{(d-1)/2}}
=-\frac{C_{\mathrm{F}}}{d-3}\gamma^i \partial_i ((x_1-x_2)^2)^{-(d-3)/2}
\end{equation}
with coefficient
\begin{equation}
C_{\mathrm{F}}=\frac{1}{\mathrm{Vol}(S^{d-1}) } \,.
\end{equation}
In the following, we set $d=4$.
Gamma matrices satisfy anti-commutation relations,
\begin{equation}
\{ \gamma^i,\gamma^j\}=-2\eta^{ij}
\end{equation}
where $\eta^{ij}$ is mostly plus as noted above. A three-dimensional Laplacian is introduced as
\begin{equation}
\nabla^2 = \eta^{ij} \partial_i \partial_j
\end{equation}
and the differential identity
\begin{equation}
\nabla^2 \frac{1}{|x|}=-4\pi \delta (x)
\end{equation}
is satisfied. Using the above equations, we can show that
\begin{equation}
\gamma^i \partial_i G_{\mathrm{F}} (x_1,x_2)=
-4\pi \delta (x_{12}) \,.
\end{equation}
We use above equations in the following computations.

We wish to evaluate one-loop corrections of propagators and vertex operators.
For the photon propagator, the one-loop correction does not diverge logarithmically because the internal fermion propagators gives an integral like
\begin{equation}
\int \! \dd^3z_1 \dd^3 z_2 \, \frac{1}{|z_{12}|^4}
\end{equation}
where $z_{12}=z_1-z_2$. This integral does not give any logarithmic divergence and it is not necessary to regulate the integral for our purpose.

\begin{figure}[t]
 \begin{minipage}[b]{0.5\hsize}
  \centering
   \begin{tikzpicture}
\draw[line width=1pt] (-2.4,0) -- (0,0) -- (2.4,0);
\node at (-2.4,-0.3) {$x_1$};
\node at ( 2.4,-0.3) {$x_2$};
\node at (-1,-0.3) {$z_1$};
\node at (1,-0.3) {$z_2$};
\draw[snake it,line width=1pt] (1,0) arc [start angle = 00, end angle = 180, radius = 1];
\end{tikzpicture}
  \subcaption{One-loop correction of the fermion propagator}
\label{figa}
 \end{minipage}
 \begin{minipage}[b]{0.5\hsize}
\centering
   \begin{tikzpicture}
\draw[line width=1pt] (130:1.8) -- (0,0) -- (230:1.8);
\draw[snake it,line width=1pt] (0,0) -- (1.8,0);
\draw[snake it,line width=1pt] (130:1) arc [start angle = 130, end angle = 230, radius = 1];
\node at (130:2.1) {$x_1$};
\node at (230:2.1) {$x_2$};
\node at (-0.4,0.9) {$z_1$};
\node at (0.2,-0.3) {$z_2$};
\node at (-0.4,-1) {$z_3$};
\node at (2.1,0) {$X$};
\end{tikzpicture}
  \subcaption{One-loop correction of the vertex operator}
\label{figb}
 \end{minipage}
\caption{Feynman diagrams}
\end{figure}
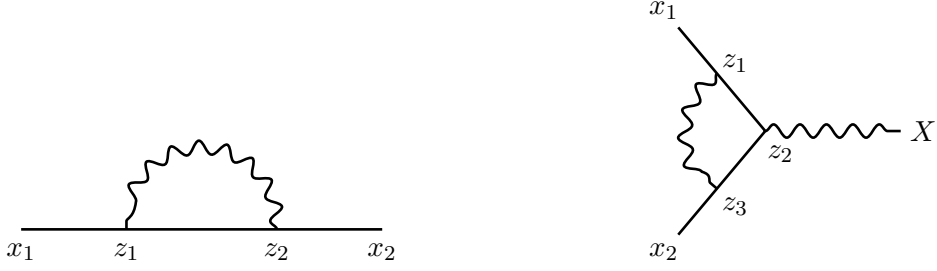

The one-loop correction of the fermion propagator (see Fig. \ref{figa}) is evaluated as
\begin{align}
\scalebox {0.5}[0.5]{
\begin{tikzpicture}[baseline={([yshift=-.5ex]current bounding box.center)}]
\draw[line width=1pt] (-2,0) -- (0,0) -- (2,0);
\draw[snake it,line width=1pt] (1,0) arc [start angle = 00, end angle = 180, radius = 1];
\end{tikzpicture}
}
&=(ig)^2 \int \! \dd^3 z_1 \dd^3 z_2 \, G_{\mathrm{F}}(x_1,z_1) \gamma^i G_{\mathrm{F}}(z_1,z_2) \gamma^j G_{\mathrm{F}}(z_2,x_2) G_{\mathrm{G},ij}(z_1,z_2) \notag \\
&=-\frac{2}{3}(ig)^2 C_{\mathrm{F}}C_{\mathrm{S}}\int \! \dd^3 z_1 \dd^3 z_2 \, G_{\mathrm{F}}(x_1,z_1)  \gamma^k \partial _k \left( \frac{1}{|z_{12}|^3}\right) G_{\mathrm{F}}(z_2,x_2) \,.
\end{align}
Note that only the photon propagator with arguments restricted to the boundary appeared in the above expressions.

The heart of differential regularization is to use the following replacement,
\begin{equation}
\frac{1}{|x|^3}=-\nabla^2 \frac{\log M|x|}{|x|}.
\label{diff-reg}
\end{equation}
This replaces the singular $|x|^{-3}$ with a much more amendable expression which, in particular, has a well defined Fourier transform. The expression \eqref{diff-reg} can easily be shown to be true for any non-vanishing $|x|$. The replacement of $|x|^{-3}$ is hence valid up to (potentially infinite) contact terms. Removing these infinite contact terms is exactly what a renormalization procedure needs to accomplish. We also see that the price to pay is that we had to introduce a mass scale $M$, just as is familiar from momentum space based regularization schemes. The $\beta$-functions of the theory will be determined by the requirement that physical quantities do not depend on the arbitrary mass scale $M$.
Using \eqref{diff-reg} we obtain
\begin{equation}
\scalebox {0.5}[0.5]{
\begin{tikzpicture}[baseline={([yshift=-.5ex]current bounding box.center)}]
\draw[line width=1pt] (-2,0) -- (0,0) -- (2,0);
\draw[snake it,line width=1pt] (1,0) arc [start angle = 00, end angle = 180, radius = 1];
\end{tikzpicture}
}=-\frac{2}{3}(ig)^2 C_{\mathrm{F}}C_{\mathrm{S}}\int \! \dd^3 z_1 \dd^3 z_2 \, G_{\mathrm{F}}(x_1,z_1)  \gamma^k \partial _k \left( -\nabla^2 \frac{\log M|z_{12}|}{|z_{12}|} \right) G_{\mathrm{F}}(z_2,x_2) \,.
\end{equation}
Eventually we get
\begin{equation}
M\frac{\partial}{\partial M} \left(
\scalebox {0.5}[0.5]{
\begin{tikzpicture}
\draw[line width=1pt] (-2,0) -- (0,0) -- (2,0);
\draw[snake it,line width=1pt] (1,0) arc [start angle = 00, end angle = 180, radius = 1];
\end{tikzpicture}
}
\right)
 =\frac{g^2}{6\pi^2}G_{\mathrm{F}}(x_1,x_2) \,.
\end{equation}

Finally, we evaluate the one-loop correction of the vertex operator (See Fig. \ref{figb}),
\begin{align}
\scalebox {0.4}[0.4]{
\begin{tikzpicture}[baseline={([yshift=-.5ex]current bounding box.center)}]
\draw[line width=1pt] (130:1.8) -- (0,0) -- (230:1.8);
\draw[snake it,line width=1pt] (0,0) -- (1.4,0);
\draw[snake it,line width=1pt] (130:1) arc [start angle = 130, end angle = 230, radius = 1];
\end{tikzpicture}
}
&=(ig)^3 \int \! \dd^3 z_1 \dd^3 z_2 \dd^3 z_3\, G_{\mathrm{F}}(x_1,z_1) \gamma^i G_{\mathrm{F}}(z_1,z_2) \gamma^j G_{\mathrm{F}}(z_2,z_3) \gamma^m \notag \\
&\hspace{40mm} \times G_{\mathrm{F}}(z_3,x_2)  G_{\mathrm{G},im}(z_1,z_3) G_{\mathrm{G},j\mu}(z_2,X) \notag \\
&=2(ig)^3 C_{\mathrm{F}}^2 C_{\mathrm{S}} \int \! \dd^3 z_1 \dd^3 z_2 \dd^3 z_3\, G_{\mathrm{F}}(x_1,z_1) \gamma^i \gamma^k \gamma^j \gamma^l \gamma _i \notag \\
&\hspace{30mm} \times \partial _k \left( \frac{1}{|z_{12}|}\right) \partial _l \left( \frac{1}{|z_{23}|}\right) \frac{1}{|z_{13}|^2} G_{\mathrm{F}}(z_3,x_2) G_{\mathrm{G},j\mu}(z_2,X) \,.
\end{align}
To pick up a logarithmic divergent part, we use a following replacement
\begin{equation}
\begin{split}
&\partial _k \left( \frac{1}{|z_{12}|}\right) \partial _l \left( \frac{1}{|z_{23}|}\right) \frac{1}{|z_{12}+z_{23}|^2}
=\partial _k \left( \frac{1}{|z_{12}|} \partial _l \left( \frac{1}{|z_{23}|}\right) \frac{1}{|z_{12}+z_{23}|^2} \right)  \\
&-\partial _k \left( \frac{1}{|z_{12}|}\right) \frac{1}{|z_{23}|} \partial _l \left( \frac{1}{|z_{12}+z_{23}|^2} \right)
-\partial _k \partial _l \left( \frac{1}{|z_{12}|}\right) \frac{1}{|z_{23}| |z_{12}+z_{23}|^2 }\,.
\end{split}
\end{equation}
Note that $\partial _k$ is a derivative in terms of $z_1$ while $\partial _l$ is a derivative in terms of $z_2$.
The first and second terms do not give any logarithmic divergences, so we ignore them and concentrate on the third term from now on.
After changing the variable from $z_2$ of $\partial _l$ to $z_1$, we get an identity,
\begin{equation}
\begin{split}
&\partial _k \partial _l \left( \frac{1}{|z_{12}|}\right) \frac{1}{|z_{23}| |z_{12}+z_{23}|^2 }  \\
&= \left(\partial _k \partial _l - \frac{1}{3} \eta_{kl} \nabla^2 \right)\left( \frac{1}{|z_{12}|}\right) \frac{1}{|z_{23}| |z_{12}+z_{23}|^2 }
+\frac{1}{3} \eta_{kl} \nabla^2 \left( \frac{1}{|z_{12}|}\right) \frac{1}{|z_{23}| |z_{12}+z_{23}|^2 }\,.
\end{split}
\end{equation}
The first term does not give any contribution when we solve Callan-Symanzik equation.
Eventually, the one-loop correction of the vertex operator reduces to
\begin{align}
&(ig)^3 \frac{2}{3} C_{\mathrm{F}}^2 C_{\mathrm{S}}
\int \! \dd^3 z_1 \dd^3 z_2 \dd^3 z_3\, G_{\mathrm{F}}(x_1,z_1) \gamma^j \nabla^2 \left( \frac{1}{|z_{12}|} \right)  \frac{1}{|z_{23}| |z_{12}+z_{23}|^2 } G_{\mathrm{F}}(z_3,x_2) G_{\mathrm{G},j\mu}(z_2,X) \notag \\
&=-(ig)^3 \frac{2\cdot 4 \pi }{3} C_{\mathrm{F}}^2 C_{\mathrm{S}} \int \! \dd^3 z_2 \dd^3 z_3\, G_{\mathrm{F}}(x_1,z_2) \gamma^j \frac{1}{|z_{23}|^3} G_{\mathrm{F}}(z_3,x_2) G_{\mathrm{G},j\mu}(z_2,X) \notag \\
&=-(ig)^3 \frac{2\cdot 4 \pi }{3} C_{\mathrm{F}}^2 C_{\mathrm{S}} \int \! \dd^3 z_2 \dd^3 z_3\, G_{\mathrm{F}}(x_1,z_2) \gamma^j \nabla^2 \left( -\frac{\log M|z_{23}| }{|z_{23}|} \right) G_{\mathrm{F}}(z_3,x_2) G_{\mathrm{G},j\mu}(z_2,X)
\end{align}
where we used
\begin{equation}
\gamma ^i \gamma ^k \gamma^j \gamma_k \gamma_i = \gamma^j \,.
\end{equation}
Thus the derivative of the one-loop correction of the vertex operator becomes
\begin{align}
M\frac{\partial }{\partial M} \left(
\scalebox {0.4}[0.4]{
\begin{tikzpicture}[baseline={([yshift=-.5ex]current bounding box.center)}]
\draw[line width=1pt] (130:1.8) -- (0,0) -- (230:1.8);
\draw[snake it,line width=1pt] (0,0) -- (1.4,0);
\draw[snake it,line width=1pt] (130:1) arc [start angle = 130, end angle = 230, radius = 1];
\end{tikzpicture}
}
\right)&= -(ig)^3 \frac{2 (4\pi)^2 }{3} C_{\mathrm{F}}^2 C_{\mathrm{S}}
\int \! \dd^3 z \, G_{\mathrm{F}}(x_1,z) \gamma^j G_{\mathrm{F}}(z,x_2) G_{\mathrm{G},j\mu}(z,X) \notag \\
&=\frac{g^2}{6\pi^2}
\times
\scalebox {0.4}[0.4]{
\begin{tikzpicture}[baseline={([yshift=-.5ex]current bounding box.center)}]
\draw[line width=1pt] (130:1.8) -- (0,0) -- (230:1.8);
\draw[snake it,line width=1pt] (0,0) -- (1.4,0);
\end{tikzpicture}
} \,.
\end{align}

To determine the corresponding $\beta$-function and anomalous dimensions, we need to solve the Callan-Symanzik equations which encode independence of physical correlation functions from the arbitrary RG scale $M$.
The Callan-Symanzik equations are given by
\begin{align}
&\left( M\frac{\partial }{\partial M} +\beta \frac{\partial}{\partial g} + 2\gamma _g \right)G_{\mathrm{G}}^{\mu \nu}(X_1,X_2)=0 \,, \\
&\left( M\frac{\partial }{\partial M} +\beta \frac{\partial}{\partial g} + 2\gamma _\psi \right)G_{\mathrm{F}}(x_1,x_2)=0\,,  \\
&\left( M\frac{\partial }{\partial M} +\beta \frac{\partial}{\partial g} + \gamma_g +2 \gamma_\psi \right)\Gamma_g (x_1,x_2,X)=0 \,.
\end{align}
The $\beta$-function and $\gamma$ can be expanded as
\begin{align}
\beta&= \beta_1 g^2+\beta_2 g^3+\mathcal{O}(g^4) \,,  \label{b1} \\
\gamma_g &= \gamma_{g,1} g^2 + \gamma_{g,2} g^3+\mathcal{O}(g^4) \,, \label{b2}  \\
\gamma_\psi &=\gamma_{\psi,1} g^2 + \gamma_{\psi,2} g^3+\mathcal{O}(g^4) \,. \label{b3}
\end{align}
Note that we can easily check that the order $\mathcal{O}(g^0)$ and $\mathcal{O}(g^1)$ vanish in $\beta$ and $\gamma$.
Substituting \eqref{b1}, \eqref{b2} and \eqref{b3} into the Callan-Symanzik equations,
we easily obtain
\begin{align}
&\beta_1=\beta_2=0 \,, \\
& \gamma_{g,1} = \gamma_{g,2} =0 \,, \\
& \gamma_{\psi,1} = - \frac{1}{6\pi^2} \,.
\end{align}
These results are in perfect agreement with \cite{HH}. While these calculation only establish conformality of mixed dimensional QED through the first few orders in perturbation theory, it was argued in \cite{HH} that this behavior will continue to all orders in the perturbative expansion.

\subsection{Mixed dimensional QED with jumping coupling constant}

To obtain non-supersymmetric conformal manifolds with both ambient and boundary deformations, we add a marginal ambient operator to mixed dimensional QED. As we saw, the kinetic term of a gauge field with a jumping coupling constant is a nice candidate. We consider the Lagrangian,
\begin{equation}
S=-\frac{1}{4} \int \! \dd ^4 X\, (1-\gamma \varepsilon(x_3)  ) F_{\mu \nu}^2 +\int \! \dd^3 x \, i \bar{\psi} \gamma^i D_i \psi
\end{equation}
where $\gamma =(g_+^2-g_-^2)/(g_+^2+g_-^2)$ as before.
This model has both a marginal ambient operator,
\begin{equation}
\gamma \mathcal{O}=\frac{\gamma}{4} \varepsilon (x_3) F_{\mu \nu}^2
\end{equation}
and a marginal boundary operator,
\begin{equation}
\bar{g} \mathcal{\hat{O}} =\bar{g} \bar{\psi} \gamma^i A_i \psi \,.
\end{equation}
Unlike super Janus, this model has independent couplings $\gamma$ and $\bar{g}$.

We can easily confirm that correlation functions with an odd number of boundary operator insertions vanish because they contain an odd number of gauge fields.
For correlation functions with an even number of boundary operators, we need to evaluate them individually.
At two loop order, a non-trivial correlation function comes from the three-point function with two boundary operators but it does not give any logarithmic divergence.

We can also confirm that the $\beta$-function vanishes by using differential regularization.
When compared with mixed dimensional QED, this model has a more complicated photon propagator away from the boundary. The scalar and hence the photon propagator still have the structure of \eqref{janusprop} with a direct and mirror charge term, but the relative weight of the two terms is different once we include a jumping coupling.
However, when we computed one-loop corrections of the fermion propagator and the vertex operator in mixed dimensional QED,
only propagators with both arguments on the boundary appeared.
When the arguments of the photon propagators are restricted to the defect, the propagators have a same form\footnote{There is a slight semantic difference. In subsection \ref{sec32} we assume that there is a boundary at $w=0$ but now we assume that there is a defect.}.
Consequently the one-loop corrections do not change after we add the jumping coupling constant and we conclude that the $\beta$-function still vanishes.

\section{Conclusion}
\label{sec4}

In this paper we discussed conformal manifolds with boundaries or defects.
Using conformal perturbation theory for the boundary operator, we obtained the following constraints,
\begin{gather}
B_{\mathcal{O} \mathcal{\hat{O}}}=0 \,, \\
\sum_\ell B_{\mathcal{O}\mathcal{\hat{O}}_\ell} \hat{C}_{\mathcal{\hat{O}_\ell} \mathcal{\hat{O}} \mathcal{\hat{O}} } =0
\end{gather}
in addition to similar constraints obtained in \cite{cm1,cm2}.
The two coupling constants are dialed independently except some examples and we assumed $g \sim \hat{g}^2$.
In this case, the two obtained constraints are sufficient up to order $\hat{g}^3$, but in other cases, we have to evaluate difficult higher-point correlation functions and would get additional constraints.

We studied three examples in section \ref{sec3}.
In subsection \ref{sec31}, we studied super Janus.
It has one ambient operator and two boundary operators but the coupling constants of boundary operators depend each other and we cannot keep a desired relation $g\sim \hat{g}^2$ for both boundary operators.
In this sense this model is an exceptional example.
As by-product, we obtained a simple example of conformal manifolds without boundary operators.
Next, we treated mixed dimensional QED in subsection \ref{sec32}.
As noted in \cite{HH}, this model is exactly conformal.
We reconfirmed this statement based on a differential regularization \cite{diff} which is a position space based regularization.
However, this model does not contain any ambient marginal operator and all boundary operator product expansion coefficients vanish. Hence this satisfies the constraints we obtained trivially.
To construct a non-trivial example, we add a jumping coupling constant to the mixed QED.
We confirmed that this model satisfies our new constraints and also has a vanishing $\beta$-function using differential regularization.

\section*{Acknowledgements}
YS would like to thank Kento Watanabe for useful discussion.
The work of AK is supported in part by the U.S.~Department of Energy under Grant No.~DE-SC0011637.
The work of YS is supported by the Grant-in-Aid for Japan Society for the Promotion of Science Fellows, No. 16J01567.

\appendix

\section{Asymptotic behavior of conformal blocks}
\label{app}

As well-known, the three-point function has the following general form,
\begin{equation}
\langle \mathcal{O}_1(x_1,w) \mathcal{\hat{O}}_2(x_2) \mathcal{\hat{O}}_3(x_3) \rangle
= \frac{f(\eta)}{
(x_{12}^2+w^2)^{\frac{\Delta_1+\hat{\Delta}_2-\hat{\Delta}_3}{2}}(x_{13}^2+w^2)^{\frac{\Delta_1+\hat{\Delta}_3-\hat{\Delta}_2}{2}} x_{23}^{\hat{\Delta}_2+\hat{\Delta}_3-\Delta_1} }
\end{equation}
In this appendix, we drive the asymptotic behavior of the conformal blocks, $f(\eta)$.
See section 2 in \cite{McAvity} for the asymptotic behavior of the conformal block of the two-point function in the boundary channel decomposition.

Let us remind the ambient operator can be expanded by boundary operators,
\begin{equation}
\mathcal{O}(x,w)=\sum_n \frac{B_{\mathcal{O}\mathcal{\hat{O}}_n}} {(2w)^{\Delta-\hat{\Delta}_n}}\mathcal{\hat{O}}_n (x)\,.
\tag{\ref{bope}}
\end{equation}
When the ambient operator approaches the boundary, $w\to 0$, the three-point function can be approximated as
\begin{align}
\langle \mathcal{O}_1(x_1,w) \mathcal{\hat{O}}_2(x_2) \mathcal{\hat{O}}_3(x_3) \rangle
&\sim \frac{B_{\mathcal{O}\mathcal{\hat{O}}_\ell} }{(2w)^{\Delta_1-\hat{\Delta}_\ell}} \langle \mathcal{\hat{O}}_\ell(x_1) \mathcal{\hat{O}}_2(x_2) \mathcal{\hat{O}}_3(x_3) \rangle \notag \\
&=\frac{B_{\mathcal{O}\mathcal{\hat{O}}_\ell} }{(2w)^{\Delta_1-\hat{\Delta}_\ell}} \cdot \frac{\hat{C}_{\mathcal{\hat{O}_\ell} \mathcal{\hat{O}}_2 \mathcal{\hat{O}}_3 }}{
x_{12}^{\hat{\Delta}_\ell+\hat{\Delta}_2-\hat{\Delta}_3}x_{13}^{\hat{\Delta}_\ell+\hat{\Delta}_3-\hat{\Delta}_2} x_{23}^{\hat{\Delta}_2+\hat{\Delta}_3-\hat{\Delta}_\ell}} \,.
\end{align}
Since descendant operators are sub-leading compared to primary operators, such operators can be ignored.

On the other hand, the three-point function itself can be approximated as,
\begin{align}
\langle \mathcal{O}_1(x_1,w) \mathcal{\hat{O}}_2(x_2) \mathcal{\hat{O}}_3(x_3) \rangle
&= \frac{f(\eta)}{
(x_{12}^2+w^2)^{\frac{\Delta_1+\hat{\Delta}_2-\hat{\Delta}_3}{2}}(x_{13}^2+w^2)^{\frac{\Delta_1+\hat{\Delta}_3-\hat{\Delta}_2}{2}} x_{23}^{\hat{\Delta}_2+\hat{\Delta}_3-\Delta_1} } \notag \\
&\sim \frac{\eta^{\frac{\hat{\Delta}_\ell-\Delta_1}{2}}}{w^{\Delta_1-\hat{\Delta}_\ell}} \cdot \frac{f(\eta)}{x_{12}^{\hat{\Delta}_\ell+\hat{\Delta}_2-\hat{\Delta}_3}x_{23}^{\hat{\Delta}_2+\hat{\Delta}_3-\hat{\Delta}_\ell}x_{31}^{\hat{\Delta}_\ell+\hat{\Delta}_1-\hat{\Delta}_2} } \,.
\end{align}
Note that $w \to 0$ corresponds to $\eta \to \infty$ since the conformal cross-ratio is given by
\begin{equation}
\eta = \frac{(x_{12}^2+w^2)(x_{13}^2+w^2) }{x_{23}^2 w^2} \,.
\end{equation}

In total, the asymptotic behavior of $f(\eta)$ is given by
\begin{equation}
f(\eta) \to
\frac{B_{\mathcal{O}_1\mathcal{\hat{O}}_\ell} \hat{C}_{\mathcal{\hat{O}_\ell} \mathcal{\hat{O}}_2 \mathcal{\hat{O}}_3 }}{2^{\Delta_1-\hat{\Delta}_\ell}}  \eta^{\frac{\Delta_1-\hat{\Delta}_\ell}{2}} \,.
\end{equation}

In section \ref{sec2}, we obtained two independent solutions of the Casimir equation.
One of them behaves as
\begin{equation}
\eta^{-\alpha} {}_2F_1 \left( \frac{-1-\alpha+\beta+d-\hat{\Delta}}{2},\frac{-\alpha+\beta+\hat{\Delta}}{2},1-\alpha+\beta ,\eta \right) \to \eta^{-\frac{\alpha+\beta+\hat{\Delta}}{2}} = \eta^{\frac{\Delta_1-\hat{\Delta}}{2}} \,.
\end{equation}
Since the intermediate expression is symmetric under the exchange of $\alpha$ and $\beta$, another solution also has the same asymptotic behavior.
We confirm that overall normalization of constants of integration can be fixed from the boundary condition on the boundary.


\bibliographystyle{JHEP}
\bibliography{draft}

\end{document}